\documentclass[a4paper,11pt]{article}
\usepackage{pos}
\usepackage{amsmath}

\newcommand{\ka}{\kappa}

\newcommand{\kk}{\frac{k_0}{\kappa}}
\newcommand{\pd}{\partial}

\newcommand{\pb}{\mathbf{p}}
\newcommand{\kb}{\mathbf{k}}
\newcommand{\op}{\omega_\pb}
\newcommand{\osp}{\omega_{S(\pb)}}
\newcommand{\ap}{a_{\pb}}
\newcommand{\apd}{a^\dagger_{\pb}}
\newcommand{\bp}{b_{\pb}}
\newcommand{\bpd}{b^\dagger_{\pb}}

\title{Complex scalar field in $\kappa$-Minkowski noncommutative spacetime}

\author*[a]{Tadeusz Adach}

\affiliation[a]{University of Wroc\l{}aw, Faculty of Physics and Astronomy,\\ pl.\ M.\ Borna 9, 50-204
Wroc\l{}aw, Poland}

\emailAdd{tadeusz.adach@uwr.edu.pl}

\abstract{We present a comparison between translation charges for several Lagrangians for the $\kappa$-deformed complex scalar field using the canonical method. The Lagrangians are shown to be related by charge conjugation and twisted cyclicity, and these relationships are reflected in their conserved charges. The Lagrangian corresponding to the results obtained in \cite{Arzano:2020jro} and \cite{Bevilacqua:2022fbz} is identified, providing an explanation for the observed loss of charge conjugation symmetry under boosts.}

\FullConference{Proceedings of the Corfu Summer Institute 2024 "School and Workshops on Elementary Particle Physics and Gravity" (CORFU2024)\
12 - 26 May, and 25 August - 27 September, 2024\\
Corfu, Greece\\}

\begin{document}
\maketitle

\section{Introduction}
In the absence of direct experimental access to Planck-scale physics, one promising approach is to study the possible low-energy remnants of quantum gravity. This is the perspective of \emph{quantum gravity phenomenology}\cite{Amelino-Camelia:2008aez}\cite{Addazi:2021xuf}, which seeks to identify observable signatures of quantum gravitational effects in regimes where a full theory of quantum gravity is not required. Within this framework, effective field theory provides a natural language for organizing such effects\cite{Burgess:2020tbq}, particularly in the limit where spacetime is approximately flat. A recurring theme across many approaches to quantum gravity is the emergence of a minimal length scale, below which the classical notion of spacetime breaks down\cite{Amelino-Camelia:2008aez}\cite{Addazi:2021xuf}\cite{Mead:1964zz}\cite{Padmanabhan:1987au}. This suggests that quantum gravity may induce deformations of the symmetries of spacetime that persist even at low energies\cite{Amelino-Camelia:2008aez}\cite{Addazi:2021xuf}\cite{Amelino-Camelia:2000cpa}. One compelling realization of this idea is doubly special relativity (DSR), in which the Poincaré algebra is modified to accommodate both the speed of light and a fundamental length or energy scale as invariants\cite{Amelino-Camelia:2000stu}\cite{Kowalski-Glikman:2001vvk}. 

A particularly well-studied deformation of this type is the noncommutative $\kappa$-Poincaré 
 Hopf algebra\cite{Lukierski:1991pn}\cite{Kowalski-Glikman:2002iba}. It has been shown to describe the deformed symmetries of so-called $\kappa$-Minkowski spacetime\cite{Majid:1994cy}\cite{Kowalski-Glikman:2002eyl} described by the commutation relations of the $\mathfrak{an}(3)$ Lie algebra:
\begin{equation}
\label{kmin}
\left[\hat{x}^0,\hat{x}^j\right]=\frac i \kappa \hat{x}^j,
\end{equation}
where the deformation parameter $1/\kappa$ corresponds to an invariant length.

In a recent series of papers\cite{Arzano:2020jro}\cite{Bevilacqua:2022fbz} a comprehensive analysis was conducted on the symmetry properties of the complex scalar field in $\kappa$-Minkowski, focusing on the discrete ($CPT$) sector. The deformed continuous symmetries were expressed in the classical basis\cite{Kosinski:1994br}\cite{Freidel:2007hk}, in which the algebra sector of $\kappa$-Poincaré remains undeformed, while the coalgebra sector becomes highly nontrivial. This choice of basis results in an undeformed mass-shell relation and Noether charges that satisfy the traditional Poincaré algebra. However, the main result of \cite{Bevilacqua:2022fbz} was that the conserved boost charges turned out not to be $C$-invariant despite the action having manifest $C$-symmetry, which carries significant phenomenological consequences. 

Due to technical difficulties, the charges used for the aforementioned analysis were obtained using the covariant phase space formalism, which led to some ambiguity in their exact form. Having overcome said challenges, the aim of this work is to compare the charges with ones obtained with the canonical method and analyze the extent and possible origins of $C$-breaking. We restrict our analysis to the description of translation charges, which suffices to illustrate the main points of our study. We postpone to future work the analysis of the charges associated with the Lorentz sector, which, due to its technical challenge, deserves detailed scrutiny.

This paper is organized as follows: in Section 2 we provide a brief summary of the mathematical tools and identities used to perform the necessary calculations. For a more detailed analysis, see \cite{Freidel:2007hk}, \cite{Arzano:2020jro}. We also introduce the two simplest actions for the complex scalar field. In Section 3, we calculate the translation charges for the two actions and explain their relation with one another. In Section 4, we repeat the analysis for a symmetrized action and compare the result with the one obtained in \cite{Bevilacqua:2022fbz} with a focus on $C$-symmetry. Finally, in Section 5, we provide a brief discussion of the findings and present an outlook for ongoing and future research.
\section{Preliminaries}
In this section we introduce the necessary mathematical tools for the analysis of fields on noncommutative spacetime.
\subsection{Momentum space and Weyl map}
Given that the spacetime coordinates in \eqref{kmin} correspond to algebra generators, plane waves are naturally $AN(3)$ group elements:
\begin{equation}
\hat{e}_k=e^{ik_j\hat{x}^j}e^{ik_0\hat{x}^0}
\end{equation}
where the wavenumbers $k$ are coordinates on the group manifold. Note that while the ordering in the above is arbitrary, it defines the convention used to assign the $k$ parameters.

It is at this point well established that the momentum space of $\kappa$-Minkowski is a submanifold of de Sitter\cite{Kowalski-Glikman:2002oyi}\cite{Freidel:2007hk}. In the classical basis, which is related to the $k$ parametrization by the following relations:
\begin{align}
\label{p0def}p_0 &=\ka\sinh\kk+\frac{\kb^2}{\ka}e^\kk,\\
\label{pjdef}p_j &=k_j e^\kk,\\
\label{p4def}p_4 &=\ka\cosh\kk-\frac{\kb^2}{\ka}e^\kk,
\end{align}
the momentum space coordinates correspond to an embedding of this curved space in 5-dimensional flat Minkowski\cite{Freidel:2007hk}. As such, they satisfy the condition
\begin{equation}
\label{ds4}
-p_0^2+\mathbf{p}^2+p_4^2=\kappa^2
\end{equation}
where $p_4>0$.
It is straightforward to show from the commutation relations \eqref{kmin} and the definitions \eqref{p0def}-\eqref{p4def} that $p$-parametrized plane waves obey the following composition rules:
\begin{align}
\label{oplus}
\hat{e}_p\hat{e}_q&=
e^{ip_j \hat{x}^j}e^{ip_0 \hat{x}^0}e^{iq_j \hat{x}^j}e^{iq_0 \hat{x}^0}\\&=e^{i\left(\frac{q_+}{\ka}p_j+q_j\right)\hat{x}^j}e^{i\left(\frac{q_+}{\ka}p_0+\frac{1}{p_+}p_jq_j+\frac{\ka}{p_+}q_0\right)\hat{x}^0}\equiv e^{i(p\oplus q)_j\hat{x}^j}e^{i(p\oplus q)_0 \hat{x}^0}\equiv\hat{e}_{p\oplus q}
\end{align}
where
$p_+=p_0+p_4$ and similarly for $q_+$. 
The composition law \eqref{oplus} gives us a natural notion of a ``deformed inverse'' (antipode) $S$:
\begin{align}
S(p_0)&=-p_0+\frac{\pb^2}{p_+}\nonumber\\
S(p_j)&=-\frac{\ka}{p_+}p_j\nonumber\\
S(p_4)&=p_4\label{spdef}
\end{align}
so that
\begin{equation}
p\oplus S(p)=S(p)\oplus p=0,
\end{equation}
that is to say
\begin{equation}
\label{conjugate}
\hat{e}^\dagger_p=\hat{e}_{S(p)}.
\end{equation}
We now define the Weyl map $\mathcal{W}$ by its action on $\hat{e}_p$:
\begin{align}
\mathcal{W}(e^{ip_j\hat{x}^j}e^{ip_0\hat{x}^0})=e^{ipx}
\end{align}
which gives rise to the star product
\begin{equation}
\label{stardef}
e_p\star e_q=\mathcal{W}\left(\hat{e}_p \hat{e}_q\right)=\mathcal{W}\left(\hat{e}_{p\oplus q}\right)=e_{p\oplus q}.
\end{equation}
\subsection{The complex scalar field and its action}
Given the noncommutative $\star$-product \eqref{stardef}, there are two natural choices for the scalar field action:
\begin{align}
\label{s1}S_1&=\frac{1}{2}\int d^4x\left[\left(\pd^\mu\phi\right)^\dagger\star\pd_\mu\phi-m^2\phi^\dagger\star\phi\right]\\
\label{s2}S_2&=\frac{1}{2}\int d^4x\left[\pd_\mu\phi\star\left(\pd^\mu\phi\right)^\dagger-m^2\phi\star\phi^\dagger\right]
\end{align}
It can be shown that both of these lead to the same (undeformed) on-shell relation $\omega_p^2=\pb^2+m^2$ (see below, \eqref{vars1}-\eqref{vars2}).
We assume the standard mode expansion of $\phi$ in 5-dimensional momentum space with the additional conditions \eqref{ds4} and $p_4>0$:
\begin{equation}
\label{modexp-raw}
\phi(x)=\int d^5p\,\delta(p_0^2-\op^2)\delta(\kappa^2+p_0^2-\pb^2-p_4^2)\theta(p_4)\tilde{\phi}(p)
\end{equation}
which can be expressed as
\begin{equation}
\label{modexp-full}
\phi(x)=\int\frac{d^3p}{2\op p_4/\ka}\tilde{\phi}(\op,\pb)e^{ipx}+\int\frac{d^3 S(p)}{2\osp p_4/\ka}\tilde{\phi}(S(\op),S(\pb))e^{iS(p)x}
\end{equation}
by taking advantage of the fact that $\op>0\iff S(\op)<0$. To simplify notation, we will express \eqref{modexp-full} as
\begin{equation}
\label{mod-fl}
\phi(x)=\int\frac{d^3p}{\sqrt{2\op p_4/\ka}}\left[\ap e^{ipx}+\bpd e^{iS(p)x}\right],
\end{equation}
which, using \eqref{conjugate}, gives us the conjugate field
\begin{equation}
\label{mod-fld}
\phi^\dagger(x)=\int\frac{d^3p}{\sqrt{2\op p_4/\ka}}\left[\apd e^{iS(p)x}+\bp e^{ipx}\right].
\end{equation}
The normalization was chosen to facilitate the comparison of results with \cite{Bevilacqua:2022fbz}.
\subsection{Differential calculus}
We define the partial derivatives by their action on plane waves:
\begin{align}
\pd_\mu e^{ipx}=ip_\mu,\\
\pd_4 e^{ipx}=i(p_4-\ka).
\end{align}

This naturally leads to deformations of Leibniz rule inherited from composition rules for $p$ through Fourier transforms:
\begin{equation}
\pd_\mu(f(x)\star g(x))=\int \mu(p) \mu(q)  f(p)g(q)\pd_\mu e^{i(p\oplus q)x},
\end{equation}
where $\mu(p)$ is shorthand for the integration measure in \eqref{modexp-raw}.
The rules can be summed up as follows:
\begin{align}
\label{pd0}
\pd_0(f(x)\star g(x))=\partial_0 (f\star g) = 
	\partial_0f \star \left(\frac{\Delta_+}{\ka}g\right)
	+ 
	\left(\frac{\ka}{\Delta_+}f\right)\star(\partial_0 g)
	+i
	(\Delta_+^{-1}\partial_j f)\star(\partial_j g),
\end{align}
\begin{align}
\label{pdj}
    \partial_j (f(x)\star g(x)) =
	(\partial_j f)\star \left(\frac{\Delta_+}{\kappa}g\right)
	+
	f\star(\partial_j g),
\end{align}
where the $\Delta_+=-i\pd_0-i\pd_4+\ka$ operator is such that $\Delta_+e^{ipx}=p_+e^{ipx}$. We also define antipode derivatives in the vein of \eqref{spdef}:
\begin{align}
S(\pd_0)=-\pd_0-i\Delta_+^{-1}\pd_j\pd_j\equiv-\pd^\dagger_0\label{spd0}
\end{align}
\begin{align}
S(\pd_j)=-\frac{\ka}{\Delta_+}\pd_j\equiv-\pd_j^\dagger\label{spdj}
\end{align}
It can be shown from \eqref{pd0} and \eqref{pdj} that the following identities are satisfied:
\begin{align}
\pd_\mu f\star g&=f\star S(\pd_\mu)g+\text{total derivative}\label{total-1}\\
f\star\pd_\mu g&=S(\pd_\mu)f\star g+\text{total derivative}\label{total-2}
\end{align}
\subsection{Charge conjugation}
Using \eqref{spd0} and \eqref{spdj}, the two actions can be written as 
\begin{align}
\label{s1-2}S_1&=-\frac{1}{2}\int d^4x\left[S(\pd^\mu)\phi^\dagger\star\pd_\mu\phi+m^2\phi^\dagger\star\phi\right]\\
\label{s2-2}S_2&=-\frac{1}{2}\int d^4x\left[\pd_\mu\phi\star S(\pd^\mu)\phi^\dagger+m^2\phi\star\phi^\dagger\right]
\end{align}
If we define the $C$-transformation in the standard way as
\begin{equation}
\label{cdef}
\phi\xrightarrow[]{C}\phi^\dagger
\end{equation}
then $S_1$ and $S_2$ are related to each other by a $C$-transformation up to a surface term on account of \eqref{total-1} and \eqref{total-2}. Given the momentum space representation \eqref{mod-fl}, \eqref{mod-fld}, the transformation \eqref{cdef} assumes the standard action in momentum space: $\ap\xrightarrow{C}\bp$, etc.
\section{Noether charges}
\subsection{Variations of the two actions}
In the canonical method, the conserved charges are computed from the variation of the action. Using the derivation rules \eqref{pd0} and \eqref{pdj} we can express the variations of $S_1$ and $S_2$ as 
\begin{align}
\label{vars1}
\delta S_{1}=-\frac{1}{2}\int d^{4}x\left[\pd_{A}\left(S(\Pi_{1}^{A})\phi^{\dagger}\star\delta\phi\right)+S(\pd_{A})\left(\delta\phi^{\dagger}\star\Pi_{1}^{A}\phi\right)\right.\nonumber\\
\left.+\left(S(\pd_\mu)S(\pd^\mu)+m^2\right)\phi^\dagger\star\delta\phi+\delta\phi^\dagger\star  \left(\pd_\mu\pd^\mu+m^2\right)\phi\right]
\end{align}

\begin{align}
\label{vars2}
\delta S_{2}=-\frac{1}{2}\int d^{4}x\left[\pd_{B}\left(\delta\phi\star S(\Pi_{2}^{B})\phi^{\dagger}\right)+S(\pd_{B})\left(\Pi_{2}^{B}\phi\star\delta\phi^{\dagger}\right)\right.\nonumber\\
+\left.\delta\phi\star\left(S(\pd_\mu)S(\pd^\mu)+m^2\right)\phi^\dagger+\left(\pd_\mu\pd^\mu+m^2\right)\phi\star\delta\phi^\dagger\right]
\end{align}
with the following differential operators :
\begin{alignat}{4}
&\Pi_{1}^{0}&&=-\ka^{-1}\left(i\pd^{0}+\Delta_+\right)S(\pd^{0})\;\;\;\;\;\;\;\;\;\;\;\;\;\;\;&&\Pi_{2}^{0}&&=\ka^{-1}\left(i\pd^{0}+\Delta_+\right)\pd^{0}\nonumber\\
&\Pi_{1}^{j}&&=-\ka^{-1}\left(i\pd^{0}+\Delta_+\right)S(\pd^{j})\;&&\Pi_{2}^{j}&&=\ka^{-1}\left(i\pd^{0}+\Delta_+\right)\pd^{j}\nonumber\\
&\Pi_{1}^{4}&&=i\ka^{-1}m^{2}\;&&\Pi_{2}^{4}&&=i\ka^{-1}\pd_{0}\pd^{0}
\end{alignat}
The equations of motion are, in both cases, undeformed:
\begin{align}
\left(\pd_\mu\pd^\mu+m^2\right)\phi&=0,\\
\left(\pd_\mu\pd^\mu+m^2\right)\phi^\dagger&=0,
\end{align}
since $S(\pd_\mu)S(\pd^\mu)=\pd_\mu\pd^\mu$.
\subsection{Translations}
For translations, we use the variation $\delta^T$:
\begin{equation}
\label{transdef}
\delta^T=\epsilon ^A\pd_A.
\end{equation}

The crucial thing to note here is that the parameter $\epsilon^A$ is itself noncommutative, which allows for the operator $\delta^T$ to satisfy Leibniz rule without violating \eqref{pd0} and \eqref{pdj}:
\begin{equation}
\label{leibniz}
\delta^T(f\star g)=(\delta^T f)\star g+f\star(\delta^Tg).
\end{equation}
The conserved currents for $S_1$ are then obtained from \eqref{vars1}:
\begin{equation}
\label{varl1}
\delta^T\mathcal{L}_1+\frac{1}{2}\pd_{A}\left(S(\Pi_{1}^{A})\phi^{\dagger}\star\delta^T\phi\right)+\frac{1}{2}S(\pd_{A})\left(\delta^T\phi^{\dagger}\star\Pi_{1}^{A}\phi\right)=0,
\end{equation}
and similarly for $S_2$, which is brought into the form of a conservation equation $\pd_A T^A_{\;B}=0$.
Using \eqref{leibniz} one can show that
\begin{align}
\label{t1}
T^0_{\;A}=-\frac{1}{2}\left[\pd_{A}S(\Pi_{1}^{0})\phi^{\dagger}\star\phi+\pd_{A}\phi^{\dagger}\star\Pi_{1}^{0}\phi\right],
\end{align}
\begin{align}
\label{t2}
T_{\,B}^{0}=\frac{1}{2}\left[\pd_{B}\phi\star S(\Pi_{2}^{0})\phi^{\dagger}+\pd_{B}\Pi_{2}^{0}\phi\star\phi^{\dagger}\right],
\end{align}
where the dummy indices follow the convention of \eqref{vars1} and \eqref{vars2} to help distinguish between $S_1$ and $S_2$. Integrating over space, we obtain the translation charges in a rather straightforward calculation:
\begin{align}
\label{p1}
\mathcal{P}_\mu ^1=\int d^3 x\,T^{0}_{\;\mu}=\frac{1}{2}\int d^3 p\left[\frac{p_+^{3}}{\ka^{3}}p_{\mu}\bpd\bp-S(p_{\mu})\apd\ap\right]
\end{align}
\begin{align}
\label{p2}
\mathcal{P}_\mu ^2=\int d^3 x\,T^{0}_{\;\mu}=\frac{1}{2}\int d^3 p\left[\frac{p_+^{3}}{\ka^{3}}p_{\mu}\apd\ap-S(p_{\mu})\bpd\bp\right]
\end{align}
It is immediately obvious that the two charges are related by a $C$-transformation, as could be expected, since the total derivative by which $S_1^C$ differs from $S_2$ is exactly what is used to calculate the charge. Note that the momentum $p_\mu$ and its antipode $S(p_\mu)$ come directly from the $\pd_A/\pd_B$ derivatives acting on the left field in \eqref{t1} and \eqref{t2} and thus can be traced back to the mode expansions \eqref{mod-fl} and \eqref{mod-fld}.
\section{The $C$-invariant action}
Since $S_1$ and $S_2$ are related by a $C$-transformation, we can construct a manifestly $C$-invariant action via
\begin{equation}
\label{sc}
S_C=\int d^4 x\mathcal{L}_C=\frac 1 2\left(S_1+S_2\right).
\end{equation}
This is the approach taken in \cite{Arzano:2020jro} and \cite{Bevilacqua:2022fbz}. The resulting charge is of the form
\begin{equation}
\label{andreacharge}
\mathcal{P}_\mu=\frac{1}{2}\int d^3 p\left[-S(p_\mu)\ap\apd+p_\mu\bp\bpd\right].
\end{equation}
We see that, despite symmetrization, the result is structurally akin to the one for the asymmetric action $S_1$ apart from the $p_+^3/\ka^3$ factor. It should be noted that \eqref{andreacharge} was obtained using the covariant phase space formalism, that is by contracting the vector field generating translations with the symplectic form (for a comprehensive modern review of the method, see \cite{Harlow:2019yfa}):
\begin{equation}
\label{cps}
-\delta^T_\mu \lrcorner \,\Omega=\delta\mathcal{P}_\mu.
\end{equation}
In the following section we will demonstrate how the same result can be obtained canonically from the variation of the action \eqref{sc}.
\subsection{Variation of the symmetric action}
The variation of \eqref{sc} is trivially the sum of variations of $S_1$ and $S_2$ (\eqref{vars1} and \eqref{vars2}). However, one notices that the equation of motion terms, e.g.
\begin{align}
-\frac 1 4 \int d^4 x\left[\delta\phi^\dagger\star  \left(\pd_\mu\pd^\mu+m^2\right)\phi+  \left(\pd_\mu\pd^\mu+m^2\right)\phi\star\delta\phi^\dagger\right]
\end{align}
do not allow for a simple factorization due to the noncommutative nature of the  $\star$-product. Under integration, the ordering problem can be resolved via \emph{twisted cyclicity} (see Appendix A). If one chooses to put $\delta\phi^\dagger$ on the left, the resulting deformed equation of motion for $\phi$ is then 
\begin{equation}
\label{eom-def}
\left(1+\frac{\Delta_+^3}{\kappa^3}\right)\left(\pd_\mu\pd^\mu+m^2\right)\phi=0.
\end{equation}
Since on-shell the quantity $p_+^3/\ka^3$ is always positive, the equations of motion shouldn't be affected and in principle the analysis can be continued, yielding
\begin{equation}
\mathcal{P}_\mu^C=\mathcal{P}_\mu^1+\mathcal{P}^2_\mu,
\end{equation}
which is manifestly $C$-symmetric. However, when this approach is extended to the Lorentz sector, the full set of charges does not seem to satisfy the Poincaré algebra. This issue is beyond the scope of this work and will be analyzed in detail in a future publication.

One can also apply twisted cyclicity a priori and consider the action $S_C$ as 
\begin{equation}
S_C=\int d^4 x\mathcal{L}_{C1}=-\frac{1}{2}\int d^4 x\left[S(\pd_\mu)\phi^\dagger\star\pd_\mu\left(1+\frac{\Delta_+^3}{\kappa^3}\right)\phi+m^2\phi^\dagger\star\left(1+\frac{\Delta_+^3}{\kappa^3}\right)\phi \right].
\end{equation}
In this form, the analysis can be carried out exactly as for $\mathcal{L}_1$, resulting in the translation charges
\begin{equation}
\label{pc1}
\mathcal{P}^{C1}_\mu=\frac{1}{4}\int d^3 p\left(1+\frac{p_+^3}{\ka^3}\right)\left[-S(p_\mu)\apd\ap+p_\mu\bpd\bp\right],
\end{equation}
which coincides with \eqref{andreacharge} (in which the overall factor $1+p_+^3/\ka^3$ was absorbed into the definitions of $\ap$ and $\bp$).

The Lagrangians $\mathcal{L}_C$ and $\mathcal{L}_{C1}$, however, are not equivalent, and since their actions are equal we can assume they differ by a total derivative, which is presumably responsible for breaking $C$-symmetry. Note that the choice to put $\phi^\dagger$ on the left was arbitrary, and one could just as well consider  the Lagrangian
\begin{align}
\mathcal{L}_{C2}=-\frac{1}{4}\left[\pd_\mu\left(1+\frac{\kappa^3}{\Delta_+^3}\right)\phi\star S(\pd^\mu)\phi^\dagger+m^2\left(1+\frac{\kappa^3}{\Delta_+^3}\right)\phi\star \phi^\dagger\right]
\end{align}
which leads to the charge
\begin{equation}
\label{pc2}
\mathcal{P}^{C2}_\mu=\frac{1}{4}\int d^3 p\left(1+\frac{p_+^3}{\ka^3}\right)\left[p_\mu\apd\ap-S(p_\mu)\bpd\bp\right],
\end{equation}
that is related by charge conjugation to \eqref{pc1}. However, $\mathcal{L}_{C1}$ and $\mathcal{L}_{C2}$ (and similarly $\mathcal{L}_1$ and $\mathcal{L}_2$) are physically equivalent, as the only way of distinguishing the two types of particles in the theory is their momentum.

\subsection{The fate of $C$-symmetry}
The only way of preserving $C$-symmetry (as defined in \eqref{cdef}) in the deformed theory appears to be the direct use of the symmetric Lagrangian $\mathcal{L}_C$. Its treatment, however, presents significant technical challenges and its conserved charges appear not to satisfy the Poincaré algebra, though it is not yet completely clear whether those problems stem from the methodology or the Lagrangian itself.  Its deeper analysis will be presented in an upcoming paper, along with new insight into the fate of discrete symmetries in $\kappa$-deformed theories. 

\section{Conclusions}
We have shown that the canonical method is in full agreement with the covariant phase space formalism in the case of translation charges for $\kappa$-deformed fields. Their lack of charge conjugation symmetry has been explained by an implicit change of Lagrangian, which appears necessary in both formalisms, suggesting a possible inherent incompatibility between $C$-symmetry and $\kappa$-deformation unless it were redefined.

Since $|S(p_\mu)|<|p_\mu|$ and $p_+/\ka>1$, the choice between $\mathcal{L}_C$, $\mathcal{L}_{C1}$/ $\mathcal{L}_{C2}$ or $\mathcal{L}_1$/$\mathcal{L}_2$ is phenomenologically meaningful. $\mathcal{L}_C$ is currently not fully understood, while between the latter two pairs there is no preference on theoretical grounds, although they could, in principle, be distinguished by experiment. It remains to be seen whether higher spin fields provide additional insight into ordering and the behavior of charge conjugation.
\section*{Acknowledgements}
The author's participation in \emph{Corfu Summer Institute 2024 "School and Workshops on Elementary Particle Physics
and Gravity" } was supported by an ITC Conference Grant from COST action CA21109.
\appendix

\section{Twisted cyclicity}
While the $\star$-product is inherently noncommutative, it possesses a property known as \emph{twisted cyclicity} under integration that allows some form of reordering:
\begin{equation}
\int d^4xf\star g=\int d^4 xg\star\left(\frac{\Delta_+^3}{\kappa^3}f\right)=\int d^4 x\left(\frac{\kappa^3}{\Delta_+^3}g\right)\star f
\end{equation}
This can be proven directly from the Fourier transform:
\begin{align*}
\int d^{4}xf(x)\star g(x)	&=\int d^{4}x\int \frac{d^{4}p}{p_4/\ka}\int \frac{d^{4}S(q)}{q_4/\ka}f(p)g(S(q))e^{i(p\oplus S(q))x}\\
	&=\int \frac{d^{4}p}{p_4/\ka}\int d^{4}S(q)f(p)g(S(q))\frac{q_{+}^{3}}{\ka^{3}}\delta^{4}(p-q)\\
	&=\int \frac{d^{4}S(q)}{q_4/\ka}f(q)g(S(q))\frac{q_{+}^{3}}{\ka^{3}}=\int \frac{d^{4}q}{q_4/\ka} f(q)g(S(q))
\end{align*}
\begin{align*}
\int d^{4}x g(x)\star f(x)	&=\int d^{4}x\int \frac{d^{4}S(p)}{p_4/\ka}\int \frac{d^{4}q}{q_4/\ka} f(q)g(S(p))e^{i(S(p)\oplus q)x}\\
	&=\int \frac{d^{4}S(p)}{p_4/\ka}\int d^{4}q f(q)g(S(p))\frac{\ka^{3}}{q_{+}^{3}}\delta^{4}(S(p)-S(q))\\
	&=\int \frac{d^{4}q}{q_4/\ka} f(q)g(S(q))\frac{\ka^{3}}{q_{+}^{3}}=\int d^{4}x\frac{\ka^{3}}{\Delta_{+}^{3}}f(x)\star g(x)\\
    &=\int d^{4}x f(x)\star\frac{\Delta_{+}^{3}}{\ka^{3}}g(x)
\end{align*}
where we used the formula $\delta^4(p\oplus S(q))=q_+^3/\ka^3q_4/\ka\delta^4(p-q)$.

\end{document}